\documentstyle[aps,floats,epsfig]{revtex}

\newcommand{\ket}[1]{\left | \, #1 \right \rangle}

\begin{document}
\draft
\twocolumn[\hsize\textwidth\columnwidth\hsize\csname
@twocolumnfalse\endcsname

\title{Observation of three-photon Greenberger-Horne-Zeilinger entanglement}
\author{Dik Bouwmeester, Jian-Wei Pan, Matthew Daniell, Harald Weinfurter \& Anton Zeilinger}
\address{Institut f\"{u}r Experimentalphysik,
Universit\"{a}t Innsbruck, Technikerstr. 25, A-6020 Innsbruck,
Austria}
\date{submitted to PRL}
\maketitle
\begin{abstract}
We present the experimental observation of polarization entanglement for three spatially separated photons. Such
states of more than two entangled particles, known as GHZ states, play a crucial role in fundamental tests of
quantum mechanics versus local realism and in many quantum information and quantum computation schemes. Our
experimental arrangement is such that we start with two pairs of entangled photons and register one photon in a way
that any information as to which pair it belongs to is erased. The registered events at the detectors for the
remaining three photons then exhibit the desired GHZ correlations.
\end{abstract}
\vskip2pc
]

\narrowtext

Ever since the seminal work of Einstein, Podolsky and Rosen \cite{EINSTEIN} there has been a quest for generating
entanglement between quantum particles. Although two-particle entanglements have long been demonstrated
experimentally \cite{WU,BBO}, the preparation of entanglement between three or more particles remains an
experimental challenge. Proposals have been made for experiments with photons \cite{ZEIL97} and atoms \cite{PARIS},
and three nuclear spins within a single molecule have been prepared such that they locally exhibit three-particle
correlations \cite{NMR1}. However, until now there has been no experiment which demonstrates the existence of
entanglement of more than two spatially separated particles. Here we report the experimental observation of
polarization entanglement of three spatially separated photons.

The original motivation to prepare three-particle entanglements stems from the observation by Greenberger, Horne
and Zeilinger (GHZ) that three-particle entanglement leads to a conflict with local realism for non-statistical
predictions of quantum mechanics \cite{GHZ}. This is in contrast to the case of Einstein, Podolsky and Rosen
experiments with two entangled particles testing Bell´s inequalities,  where the conflict only arises for
statistical predictions \cite{BELL}. We will experimentally  address this issue in a forthcoming paper.

The incentive to produce GHZ states has been significantly increased by the advance of the field of quantum
communication and quantum information processing. Entanglement between several particles is the most important
feature of many such quantum communication and computation protocols \cite{A,B}.

We now describe the experimental arrangements for the observation of three-photon GHZ entanglement. The
experimental techniques are similar to those that have been developed for our previous experiments on quantum
teleportation \cite{BOU97} and entanglement swapping \cite{PAN98}. In fact, one of the main complications in the
previous experiments, namely the creation of two pairs of photons by a single source, is here turned into a virtue.

The main idea, as was put forward in Ref.\cite{ZEIL97}, is to transform two pairs of polarization entangled photons
into three entangled photons and a fourth independent photon. Figure 1 is a schematic drawing of our experimental
setup. Pairs of polarization entangled photons are generated by a short (approx. 200 fs) pulse of UV-light of
wavelength $\lambda=788nm$ from a mode-locked Ti-Sapphire laser, which passes through an optically nonlinear
crystal (here Beta-Barium-Borate, BBO). The probability per pulse to create a single pair in the desired modes is
rather low and of the order of a few $10^{-4}$. The pair creation is such that the following polarization entangled
state is obtained \cite{BBO}.
\begin{equation}
\label{polstate}
\frac{1}{\sqrt{2}} \left( \ket{H}_a\ket{V}_b - \ket{V}_a\ket{H}_b \right) \,.
\end{equation}
This state indicates that there is a superposition of the possibility that the photon in arm $a$ is horizontally
polarized and the one in arm $b$ vertically polarized ($\ket{H}_a\ket{V}_b$), and the opposite possibility, i.e.,
$\ket{V}_a\ket{H}_b$. The minus sign indicates that there is a fixed phase difference of $\pi$ between the  two
possibilities. For our GHZ experiment this phase factor is actually allowed to have any value, as long as it is
fixed for all pair creations.

The setup is such that arm $a$ continues towards a polarizing beamsplitter, where $H$ photons are transmitted
towards detector T and $V$ photons are reflected, and arm $b$ continues towards a 50/50 polarization-independent
beamsplitter. From each beamsplitter one output is directed to a second polarizing beamsplitter. In between the two
polarizing beamsplitters there is a $\lambda/2$ retardation plate at an angle of $22.5^{\circ}$ which rotates the
vertical polarization of the photons reflected by the first polarizing beamsplitter into a $45^{\circ}$
polarization, i.e. a superposition of $\ket{H}$ and $\ket{V}$ with equal amplitudes. We use three more detectors,
$D_1$, $D_2$, and $D_3$, in the remaining output arms.  Narrow band-width interference filters
are placed in front
of the four detectors ($\delta\lambda=4.5\,$ nm for the detector T and $\delta\lambda=3.6\,$ nm for the other three).
Including filter losses, coupling into single-mode fibers, and the Si-avalanche detector
efficiency, the total collection and detection probability of a photon is about 10\%.

Consider now the case that {\em two} pairs are generated by a
single UV-pulse, and that the four photons are all detected, one by
each detector T, $D_1$, $D_2$, and $D_3$. Our claim is that by the
coincident detection of four photons and because of the brief
duration of the UV pulse and the narrowness of the filters, one can
conclude that a three-photon GHZ state has been recorded by
detectors D$_1$, D$_2$, and D$_3$. The reasoning is as follows.
When a four-fold coincidence recording is obtained, one photon in
path $a$ must have been horizontally polarized and detected by the
trigger detector T. Its companion photon in path $b$ must then be
vertically polarized, and it has 50\% chance to be transmitted by
the beamsplitter (see Figure 1) towards detector D$_3$ and 50\%
chance to be reflected by the beamsplitter towards the final
polarizing beamsplitter where it will be reflected to D$_2$.
Consider the first possibility, i.e. the companion of the photon
detected at T is detected by D$_3$ and necessarily carried
polarization $V$. Then the counts at detectors D$_1$ and D$_2$ were
due to a second pair, one photon travelling via path $a$ and the
other one via path $b$. The photon travelling via path $a$ must
necessarily be $V$ polarized in order to be reflected by the
polarizing beamsplitter in path $a$; thus its companion, taking
path $b$, must be $H$ polarized and after reflection at the
beamspliter in path $b$ (with a 50\% probability) it will be
transmitted by the final polarizing beamsplitter and arrive at
detector D$_1$. The photon detected by D$_2$ therefore must be $H$
polarized since it came via path $a$ and had to transit the last
polarizing beamsplitter. Note that this latter photon was $V$
polarized but after passing the $\lambda/2$ plate it became
polarized at $45^{\circ}$ which gave it 50\% chance to arrive as an
$H$ polarized photon at detector D$_2$. Thus we conclude that if
the photon detected by D$_3$ is the companion of the T photon, then
the coincidence detection by D$_1$, D$_2$, and D$_3$ corresponds to
the detection of the state
\begin{equation}
\label{term1}
\ket{H}_1 \ket{H}_2 \ket{V}_3\,.
\end{equation}
By a similar argument one can show that if the photon detected by D$_2$ is the companion of the T photon, the
coincidence detection by D$_1$, D$_2$, and D$_3$ corresponds to the detection of the state
\begin{equation}
\label{term2}
\ket{V}_1 \ket{V}_2 \ket{H}_3\,.
\end{equation}

\begin{figure} [t]
\label{setup}
\begin{center}
\epsfig{file=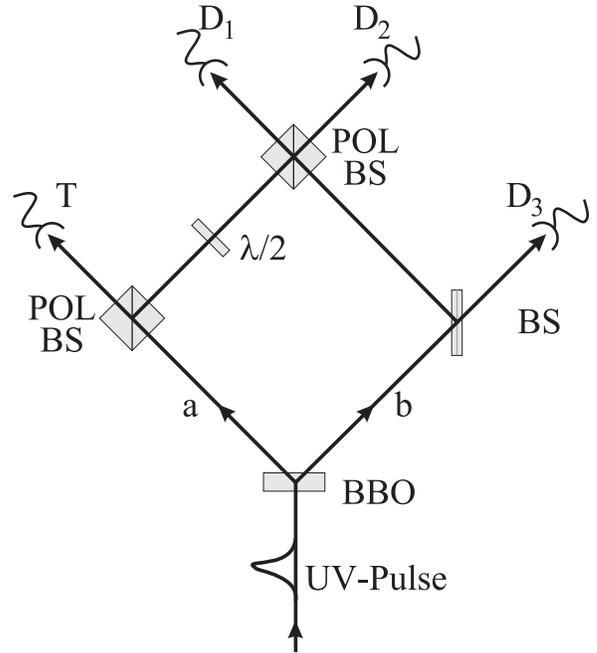,width=7.7cm}
\end{center}
\caption{Schematic drawing of the experimental setup for the
demonstration of Greenberger-Horne-Zeilinger entanglement for spatially separated photons. Conditioned on the
registration of one photon at the trigger detector T, the three photons registered at D$_1$, D$_2$, and D$_3$
exhibit the desired GHZ correlations.}
\end{figure}

In general the two possible states (\ref{term1}) and (\ref{term2})
corresponding to a four-fold coincidence recording will not form a
coherent superposition, i.e. a GHZ state, because they could, in
principle, be distinguishable. Besides possible lack of mode
overlap at the detectors, the exact detection time of each photon
can reveal which state is present. For example, state (\ref{term1})
is identified by noting that T and D$_3$, or D$_1$ and D$_2$, fire
nearly simultaneously. To erase this information it is necessary
that the coherence time of the photons is substantially longer than
the duration of the UV pulse (approx. 200 fs) \cite{MAREK}. We
achieved this by detecting the photons behind narrow band-width
filters which yield a coherence time of approx. 500 fs. Thus, the
possibility to distinguish between states (\ref{term1}) and
(\ref{term2}) is no longer present, and, by a basic rule of quantum
mechanics, the state detected by a coincidence recording of D$_1$,
D$_2$, and D$_3$, conditioned on the trigger T, is the quantum
superposition
\begin{equation}
\label{expGHZ}
\frac{1}{\sqrt{2}} \left( \ket{H}_1 \ket{H}_2 \ket{V}_3  + \ket{V}_1 \ket{V}_2 \ket{H}_3  \right) \,,
\end{equation}
which is a GHZ state \cite{REM}.

The plus sign in Eq.~(\ref{expGHZ}) follows from the following more formal derivation. Consider two
down-conversions producing the product state
\begin{equation}
\label{productstate}
\frac{1}{2} \left( \ket{H}_a\ket{V}_b-\ket{V}_a\ket{H}_b \right) \left( \ket{H}^{\prime}_a\ket{V}^{\prime}_b-
\ket{V}^{\prime}_a \ket{H}^{\prime}_b \right) \,.
\end{equation}
Initially we assume that the components $\ket{H}_{a,b}$ and $\ket{V}_{a,b}$ created in one down-conversion might be
distinguishable from the components $\ket{H}^{\prime}_{a,b}$ and $\ket{V}^{\prime}_{a,b}$ created in the other one.
The evolution of the individual components of state (\ref{productstate}) through the apparatus towards the
detectors T, D$_1$, D$_2$, and D$_3$ is given by
\begin{eqnarray}
\ket{H}_a &\rightarrow& \ket{H}_T \,,  \\
\ket{V}_b &\rightarrow& \frac{1}{\sqrt{2}}(\ket{V}_2+\ket{V}_3) \,,\\
\ket{V}_a &\rightarrow& \frac{1}{\sqrt{2}}(\ket{V}_1+\ket{H}_2) \,, \\
\ket{H}_b &\rightarrow& \frac{1}{\sqrt{2}}(\ket{H}_1+\ket{H}_3) \,.
\end{eqnarray}
Identical expressions hold for the primed components. Inserting these expressions into state
(\ref{productstate}) and restricting ourselves to those terms where only one photon is found in each output we obtain
\begin{eqnarray}
\label{stateC}
- \frac{1}{4\sqrt{2}} \left\{ \ket{H}_T \left( \ket{V}^{\prime}_1\ket{V}_2\ket{H}^{\prime}_3+
\ket{H}^{\prime}_1\ket{H}^{\prime}_2\ket{V}_3 \right) \right. \nonumber \\ \left.
+ \ket{H}^{\prime}_T \left(
\ket{V}_1\ket{V}^{\prime}_2\ket{H}_3+\ket{H}_1\ket{H}_2\ket{V}^{\prime}_3
\right) \right\} \,.
\end{eqnarray}
If now the experiment is performed such that the photon states from the two down-conversions are indistinguishable,
we finally obtain the desired state (up to an overall minus sign)
\begin{equation}
\label{expGHZfinal}
\frac{1}{\sqrt{2}}\ket{H}_T \left( \ket{H}_1 \ket{H}_2 \ket{V}_3  + \ket{V}_1 \ket{V}_2 \ket{H}_3 \right) \,.
\end{equation}

Note that the total photon state produced by our setup, i.e., the state before detection, also contains terms in
which, for example, two photons enter the same detector. In addition, the total state contains contributions from
single down-conversions. The four-fold coincidence detection acts as a projection measurement onto the desired GHZ
state (\ref{expGHZfinal}) and filters out these undesireable terms. The efficiency for one UV pump pulse to yield
such a four-fold coincidence detection is very low (of the order of $10^{-10}$). Fortunately, $7.6\times 10^{7}$
UV-pulses are generated per second, which yields about one double pair creation and detection per 150 seconds,
which is just enough to perform our experiments. Triple pair creations can be completely neglected since they can
give rise to a four-fold coincidence detection only about once every day.

To experimentally demonstrate  that a GHZ state has been obtained by the method described above, we first verified
that, conditioned on a photon detection by the trigger T, both the H$_1$H$_2$V$_3$  and the V$_1$V$_2$H$_3$
component are present and no others.  This was done by comparing the count rates of the eight possible combinations
of polarization measurements, H$_1$H$_2$H$_3$, H$_1$H$_2$V$_3$, ..., V$_1$V$_2$V$_3$. The observed intensity ratio
between the desired and undesired states was 12:1. Existence of the two terms as just demonstrated is a necessary
but not yet sufficient condition for demonstrating GHZ entanglement.  In fact, there could in principle be just a
statistical mixture of those two states. Therefore, one has to prove that the two terms coherently superpose. This
we did by a measurement of linear polarization of photon 1 along $+45^{\circ}$, bisecting the H and V direction.
Such a measurement projects photon 1 into the superposition
\begin{equation}
\label{photon1}
\ket{+45^{\circ}}_1=\frac{1}{\sqrt{2}}(\ket{H}_1 + \ket{V}_1)\,,
\end{equation}
what implies that the state (\ref{expGHZfinal}) is projected into
\begin{equation}
\label{totalstate}
\frac{1}{\sqrt{2}} \ket{H}_T\ket{+45^{\circ}}_1(\ket{H}_2 \ket{V}_3 + \ket{V}_2\ket{H}_3)\,.
\end{equation}

Thus photon 2 and 3 end up entangled as predicted under the notion of "entangled entanglement" \cite{entent}.  We
conclude that demonstrating the entanglement between photon 2 and 3 confirms the coherent superposition in state
(\ref{expGHZfinal}) and thus existence of the GHZ entanglement. In order to proceed to our experimental
demonstration we represent the entangled state (2-3) in a linear basis rotated by $45^{\circ}$. The state then
becomes
\begin{equation}
\label{45basis}
\frac{1}{\sqrt{2}} \left( \ket{+45^{\circ}}_2\ket{+45^{\circ}}_3  - \ket{-45^{\circ}}_2 \ket{-45^{\circ}}_3  \right) \,,
\end{equation}
which implies that if photon 2 is found to be polarized along -$45^{\circ}$, photon 3 is also polarized along the
same direction. We test this prediction in our experiment. The absence of the terms
$\ket{+45^{\circ}}_2\ket{-45^{\circ}}_3$ and $\ket{-45^{\circ}}_2\ket{+45^{\circ}}_3$ is due to destructive
interference and thus indicates the desired coherent superposition of the terms in the GHZ state
(\ref{expGHZfinal}). The experiment therefore consisted of measuring four-fold coincidences between the detector T,
detector 1 behind a +$45^{\circ}$ polarizer, detector 2 behind a
-$45^{\circ}$ polarizer, and measuring photon 3 behind either a +$45^{\circ}$ polarizer or a -$45^{\circ}$ polarizer.
In the experiment, the difference of arrival time of the photons at
the final polarizer, or more specifically, at the detectors D1 and
D2, was varied.

\begin{figure}
\label{result}
\begin{center}
\epsfig{file=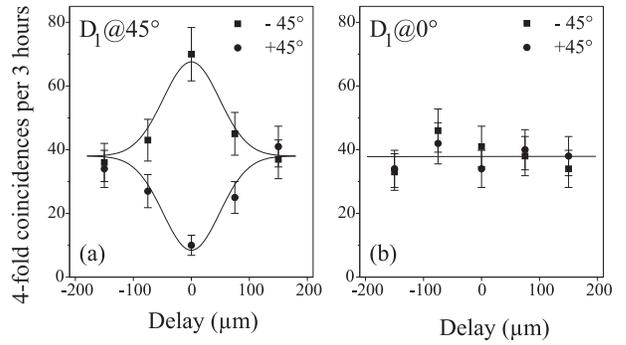,width=8cm}
\end{center}
\caption{Experimental confirmation of GHZ entanglement.
Graph (a) shows the results obtained for polarization analysis of the photon at D$_3$, conditioned on the trigger
and the detection of one photon at D$_1$ polarized at $45^{\circ}$ and one photon at detector D$_2$ polarized
$-45^{\circ}$. The two curves show the four-fold coincidences for a polarizer oriented at $-45^{\circ}$ and
$45^{\circ}$ respectively in front of detector D$_3$ as function of the spatial delay in path $a$. The difference
between the two curves at zero delay confirms the GHZ entanglement. By comparison (graph (b)) no such intensity
difference is predicted if the polarizer in front of detector D$_1$ is set at 0$^{\circ}$. }
\end{figure}

The data points in Fig.2(a) are the experimental results obtained for the polarization analysis of the photon at
D$_3$, conditioned on the trigger and the detection of two photons polarized at $45^{\circ}$ and $-45^{\circ}$ by
the two detectors D$_1$ and D$_2$, respectively. The two curves show the four-fold coincidences for a polarizer
oriented at $-45^{\circ}$ (squares) and $+45^{\circ}$ (circles) in front of detector D$_3$ as function of the
spatial delay in path $a$. From the two curves it follows that for zero delay the polarization of the photon at
D$_3$ is oriented along $-45^{\circ}$, in accordance with the quantum-mechanical predictions for the GHZ state. For
non-zero delay, the photons travelling via path $a$ towards the second polarizing beamsplitter and those travelling
via path $b$  become distinguishable. Therefore increasing the delay gradually destroys the quantum superposition
in the three-particle state.

Note that one can equally well conclude from the data that at zero delay, the photons at D$_1$ and D$_3$ have been
projected onto a two-particle entangled state by the projection of the photon at D$_2$ onto $-45^{\circ}$. The two
conclusions are only compatible for a genuine GHZ state. We note that the observed visibility was as high as 75\%.

For an additional confirmation of state (\ref{expGHZfinal}) we performed measurements  conditioned on the detection
of the photon at D$_1$  under $0^{\circ}$ polarization (i.e. $V$ polarization). For the GHZ state
(1/$\sqrt{2}$)(H$_1$H$_2$V$_3$ + V$_1$V$_2$H$_3$) this implies that the remaining two photons should be in the
state V$_2$H$_3$ which cannot give rise to any correlation between these two photons in the $45^{\circ}$ detection
basis. The experimental results of these measurement are presented in Fig.2(b). The data clearly indicate the
absence of two-photon correlations and thereby confirm our claim of the observation of  GHZ entanglement between
three spatially separated photons.

Although the extension from two to three entangled particles might seem to be only a modest step forward, the
implications are rather profound. First of all, GHZ entanglements allow for novel tests of quantum mechanics versus
local realistic models. Secondly, three-particle GHZ states might find a direct application, for example, in
third-man quantum cryptography. And thirdly, the method developed to obtain three-particle entanglement from a
source of pairs of entangled particles can be extended to obtain entanglement between many more particles
\cite{KNIGHT}, which are at the basis of many quantum communication and computation protocols. Most applications of
GHZ states imply that the three particles have to be detected. Therefore, even as our setup only produces GHZ
entanglement upon the condition that the three photons and the trigger photon are actually detected, our scheme can
readily be used for many applications. The detection plays the double role of projecting onto the GHZ state and of
performing a specific measurement on this state. Finally, we note that our experiment, together with our earlier
realization of quantum teleportation \cite{BOU97} and entanglement swapping \cite{PAN98} provides necessary to
tools to implement a number of novel entanglement distribution and network ideas as recently proposed \cite{E,F}.

We are very grateful to M.A.~Horne and D.M.~Greenberger for their useful criticism and detailed suggestions for
improvements of our initial manuscript. This work was supported by the Austrian Science Foundation FWF (Project No.
S6502), the Austrian Academy of Sciences, the US National Science Foundation NSF (Grant No. PHY 97-22614) and the
TMR program of the European Union (Network Contract No. ERBFMRXCT96-0087).



\end{document}